\documentclass[amsmath,amssymb,showkeys,superscriptaddress,aps,prd,10pt,twocolumn]{revtex4}
\usepackage{graphicx}
\usepackage[utf8]{inputenc}
\usepackage{pdfpages}
\usepackage[caption=false]{subfig}
\usepackage{multirow}
\usepackage{appendix}
\usepackage{graphicx,epstopdf}

\def\xslash{x\!\!\!\slash }

\usepackage{colortbl}
\usepackage{xcolor}
\usepackage[colorlinks=true, urlcolor=blue, linkcolor=blue, citecolor=blue]{hyperref}

\begin{document}

\title{Magnetic dipole moments of the $T_{cc}^+$ and $Z_V^{++}$ tetraquark states}

\author{K. Azizi}
\email[]{kazem.azizi@ut.ac.ir}
\affiliation{Department of Physics, University of Tehran, North Karegar Avenue, Tehran
14395-547, Iran}
\affiliation{Department of Physics, Dogus University, Acibadem-Kadikoy, 34722 
Istanbul, Turkey}
\author{U.~\"{O}zdem}%
\email[]{ulasozdem@aydin.edu.tr}
\affiliation{ Health Services Vocational School of Higher Education, Istanbul Aydin University, Sefakoy-Kucukcekmece, 34295 Istanbul, Turkey}

\date{\today}
 
\begin{abstract}

Inspired by the observation of the doubly charmed state $T_{cc}$ and  following theoretical studies on its spectroscopic parameters, we investigate its magnetic dipole moment  assigning  it the quantum numbers $J^P=1^+$ and both the compact diquark-antidiquark and molecular structures  in the framework of the light-cone QCD. We also calculate the magnetic dipole moment of the theoretically predicted singly charmed state, $Z_V^{++}$, with two units of electric charge and the quantum numbers $J^P=1^-$ in  diquark-antidiquark picture. The numerical results are obtained as $\mu_{T_{cc}^+-Di} =0.66^{+0.34}_{-0.23}~\mu_N$, $ \mu_{T_{cc}^+-Mol} =0.43^{+0.23}_{-0.22}~\mu_N $  and $\mu_{Z_{V}^{++}} =3.35^{+0.89}_{-0.73} ~\mu_N$.  These results may be checked via other phenomenological approaches. The obtained results may be useful in exact determinations of the natures of these states.

\end{abstract}

\maketitle

\section{Introduction}

Besides the standard mesons and baryons/antibaryons,  formed of $q \bar q  $ and $ qqq $/$ \bar q\bar q\bar q $, there  exist hadrons made of more than three quarks/antiquarks or valence gluons. 
These categories of hadrons  contain exotic  states including glueballs, hybrids, tetraquarks, pentaquarks, hexaquarks, etc. 
In last two decades, a large number of these exotic states have been observed in the particle factories. Although some of them are now well established, there are some doubts on existence of some other members. Exact determinations of the  nature, structure and quantum numbers of these states need more experimental efforts. Following the experiments, and sometimes before the experimental results,  these states are investigated by many theoretical and phenomenological models and approaches. There are many exotic states, proposed in theory, waiting  to be confirmed by the experiments. 
For the recent experimental and theoretical progresses on the exotic states see, for instance,  Refs.~\cite{Faccini:2012pj,Esposito:2014rxa,Chen:2016qju,Ali:2017jda,Esposito:2016noz,Olsen:2017bmm,Lebed:2016hpi,Guo:2017jvc,Nielsen:2009uh,Brambilla:2019esw,Liu:2019zoy, Agaev:2020zad, Dong:2021juy}.
Let us make a note here that most of heavy multi-quark particles discovered in the experiment, to date, have hidden-charm or hidden-bottom quark structure, i.e., they contain $ c \bar c $ or  $ b \bar b $ in their inner structures.

Very recently, the LHCb Collaboration reported observation of a first doubly charmed tetraquark state $ T_{cc}^+$  in the $D^0 D^0 \pi^+$ mass spectrum with   over 10$\sigma$ significance \cite{LHCb:2021vvq,LHCb:2021auc}.  The simplest assumption on its valence quark component is $c c\bar u \bar d$ and its spin-parity was suggested by the experiment  as $J^P = 1^+$. Its mass,  with respect to the $D^0  D^{\ast +}$ threshold,  and width have been measured to be 
\begin{eqnarray} \label{eq.Exp}
\delta m&=m_{T_{cc}^+}-(m_{D^0}+m_{D^{\ast +}})\nonumber\\
                   &=-273 \pm 61 \pm 5_{-14}^{+11}~\mathrm{KeV}, \\
\Gamma &=410 \pm 165 \pm 43_{-38}^{+18}~\mathrm{KeV}. 
\end{eqnarray}%
As we  see, its  decay width is very small and its decay to $D^0 D^0 \pi^+$ occur very slowly and with delay. Because of that this particle is the  longest living exotic state discovered till now. The discovery of the first doubly charmed tetraquark state will, undoubtedly, usher in a new era in the study of hadron spectroscopy and improve our understanding of the non-perturbative nature of the strong  interaction. Before and after the experimental discovery,  some spectroscopic properties and decay channels of the $ T_{cc}^+$ state have been investigated  within different theoretical models~ \cite{Carames:2011zz,Richard:2018yrm, Hernandez:2019eox, Liu:2019stu,Agaev:2021vur,Li:2021zbw,Yan:2021wdl,Dong:2021bvy,Feijoo:2021ppq,Meng:2021jnw,Ling:2021bir,Xin:2021wcr}.  Note that the spectroscopic parameters and different decay modes of the scalar and pseudoscalar $ T_{cc}$ states as well as the $ T_{bc}$  and $ T_{bb}$ states of different quantum numbers were already investigated in Refs. \cite{Agaev:2020mqq,Agaev:2020zag,Agaev:2020dba,Agaev:2019lwh,Agaev:2019kkz,Agaev:2019qqn,Agaev:2018khe}, which may be in agenda of future experiments.
Another important subclass of doubly charmed tetraquarks includes particles that bear two units of electric charge. Spectroscopic parameters and decay widths of such exotic states with  $cc\bar d \bar s$ and $cc\bar s \bar s$ quark contents  were  investigated in Ref. \cite{Agaev:2018vag}. These particles have not been observed experimentally yet, but their existence is important for understanding the nature of exotic states. We calculate the magnetic dipole moment of  $ T_{cc}^+$ state newly detected by LHCb collaboration \cite{LHCb:2021vvq,LHCb:2021auc} in order to shed light on its nature and physical properties.

Recently, the LHCb collaboration announced observation of two new 
structures $X_{0}(2900)$ and $X_{1}(2900)$  in the process $B^{+}\rightarrow
D^{+}D^{-}K^{+}$ \cite{Aaij:2020hon,Aaij:2020ypa}. These resonances  appear as the  intermediate states of the decay
chain $B^{+}\rightarrow D^{+}X\rightarrow D^{+}D^{-}K^{+}$, and they are neutral
four-quark states. These decays take place due to color-favored and color-suppressed
 modes of the $B^{+}$ meson, as also stated in Ref.  \cite{Burns:2020xne}. But weak decays
of $B^{+}$  may provoke also  the decays $
B^{+}\rightarrow D^{-}Z^{++}\rightarrow D^{-}D^{+}K^{+}$, where $Z^{++}$ is
a doubly charged exotic state with  inner structure of $cu\overline{s}\overline{d}$ (for more information see \cite{Agaev:2021jsz}).  It is expected that the  LHCb collaboration  observe  the  doubly charged tetraquarks $Z^{++}=[cu][
\overline{s}\overline{d}]$ with different spin-parities in the above processes.  Thus, the possible scalar
and vector four-quark resonances $Z_{\mathrm{S}}^{++}$ and $Z_{\mathrm{V}}^{++}$ may
be found as resonances in the $D^{+}K^{+}$ invariant mass.
Motivated by this, such states are proposed in Ref. \cite{Agaev:2021jsz} and the spectroscopic properties and possible decay channels of  $Z_{\mathrm{V}}^{++}$ state are  investigated using two-point QCD sum rules. Because of the importance of these states, in the second part of this study, we investigate the electromagnetic properties of $Z_{\mathrm{V}}^{++}$ state. Because of two units of electric charge, the magnetic dipole moment of this state is expected to be large and easily accessible in the future experiments.  Note that the the magnetic dipole moment of the charmed-strange $ Z_{cs} (3985)$ tetraquark was calculated in Ref. \cite{Ozdem:2021yvo} using the light-cone QCD sum rule formalism.

%

 The electromagnetic multipole moments of hadrons, besides their spectroscopic parameters, can help us determine their exact nature, substructure, and quantum numbers. These parameters give useful information about the charge and magnetization distributions inside the hadrons that help us gain information on   their geometric shapes. 
 Recall that the magnetic and higher order moments of particles that  contain information about the spatial distributions of charge and magnetization within the particles,  are directly related to the spatial distributions of quarks and gluons in hadrons.

In the present study, we calculate the magnetic dipole moments (MDM) of  $ T_{cc}^+$ and $ Z_{V}^{++}$  states  in the compact diquark-antidiquark picture using the light-cone QCD sum rule formalism. For the case of $ T_{cc}^+$, we also extract its magnetic dipole moment considering it as the $ D^*D + DD^* $ molecular state. We employ the distribution amplitudes (DAs) of the on-shell real photon state at light cone, which are available in terms of different twists. 

%
 The paper is organized in the following way: In sec. II, we calculate the MDM of the $ T_{cc}^+$ and $ Z_{V}^{++}$ states. In section III, we numerically analyze the MDM of the $ T_{cc}^+$ and $ Z_{V}^{++}$  states. The last section is devoted to
the summary of the results and conclusions. 
 
 \begin{widetext}

 \section{Formalism}
 
 In the light-cone QCD sum rule method we calculate a correlation function, as building block of the method,  once  in terms of  hadronic parameters such as electromagnetic form factors, and second, in terms  the QCD parameters and DAs of the on-shell photon. By matching the coefficients of suitable Lorentz structures from both sides and using the assumption of quark-hadron duality, we will be able to evaluate the related hadronic observables in terms of  QCD degrees of freedom as well as auxiliary parameters entering the calculations at different steps.
 
\subsection{MDM of the \texorpdfstring{$T_{cc}^+$}{} state}

To evaluate the MDM of the $T_{cc}^+$ state
 within the LCSR, we start with the correlation function 
\begin{equation}
 \label{edmn01}
\Pi _{\mu \nu }^{T_{cc}^+}(p,q)=i\int d^{4}xe^{ip\cdot x}\langle 0|\mathcal{T}\{J_{\mu}^{T_{cc}^+}(x)
J_{\nu }^{T_{cc}^+\dagger }(0)\}|0\rangle_{\gamma}, 
\end{equation}%
where  the sub-index  $\gamma$ represents the background electromagnetic field and $J_{\mu}(x)$ is the interpolating current of the $T_{cc}^+$ state
with the quantum numbers $ J^{P} = 1^{+}$ and the quark content  $ c c \bar u\bar d $.  The $T_{cc}^+$ state can be interpolated by either compact axial vector diquark-light scalar antidiquark  or molecular $ D^*D + DD^* $ structures as 
\begin{eqnarray}\label{curr}
 J_{\mu}^{T_{cc}^+-Di}(x)&=& c_a^T(x) \gamma_{\mu}C c_b(x)\bar u_a(x) C\gamma_5 \bar d_b^T(x),\\ \nonumber
  J_{\mu}^{T_{cc}^+-Mol}(x)&=& \frac{1}{\sqrt{2}}\lbrace[\bar u_a(x) i\gamma_5 c_a(x)][\bar d_b(x) \gamma_\mu c_b(x)]+[\bar u_a(x) \gamma_\mu c_a(x)][\bar d_b(x)  i\gamma_5 c_b(x)]\rbrace.
\end{eqnarray}
In Eq. (\ref{curr}), $u(x)$, $c(x)$ and $d(x)$ are the quark fields, and $C$ stands for the charge-conjugation operator. In this equation the currents $J_{\mu}^{T_{cc}^+-Di}(x)  $ and $ J_{\mu}^{T_{cc}^+-Mol}(x)  $ stand for compact diquark-antidiquark and molecular structures, respectively. In the following, we will explicitly present the calculations only for $J_{\mu}^{T_{cc}^+-Di}(x)  $ case and for simplicity we will omit the label $ Di $ from the current. However, we do all the calculations in this subsection in both pictures and will extract the numerical values of magnetic dipole moment of $T_{cc}^+$   in both pictures. 


 In hadronic side,  the correlation function is evaluated by its saturation with the intermediate hadronic states. By performing the four-integral over $x$  we get
\begin{align}
\label{edmn04}
\Pi_{\mu\nu}^{Had-T_{cc}^+} (p,q) = {\frac{\langle 0 \mid J_\mu^{T_{cc}^+} \mid
T_{cc}^+(p) \rangle}{p^2 - m_{T_{cc}^+}^2}} \langle T_{cc}^+(p) \mid T_{cc}^+(p+q) \rangle_\gamma
\frac{\langle T_{cc}^+(p+q) \mid {J^\dagger}_\nu^{T_{cc}^+} \mid 0 \rangle}{(p+q)^2 - m_{T_{cc}^+}^2} + \cdots,
\end{align}
where  q is the photon's momentum and dots stand for the contributions coming from the higher states and
continuum. The matrix element
$\langle 0 \mid J_\mu^{T_{cc}^+} \mid T_{cc}^+ \rangle$ is parameterized as
\begin{align}
\label{edmn05}
\langle 0 \mid J_\mu^{T_{cc}^+} \mid T_{cc}^+ \rangle = \lambda_{T_{cc}^+} \varepsilon_\mu^\theta\,,
\end{align}
with $\lambda_{T_{cc}^+}$ being the residue of the $T_{cc}^+$ state and $ \varepsilon_\mu^\theta\ $ is its polarization vector.

In the existence of the external electromagnetic background field, the matrix element $\langle T_{cc}^+(p) \mid  T_{cc}^+ (p+q)\rangle_\gamma $ can be expressed  in terms of  the Lorentz invariant form factors as follows~\cite{Brodsky:1992px}:
\begin{align}
\label{edmn06}
\langle T_{cc}^+(p,\varepsilon^\theta) \mid  T_{cc}^+ (p+q,\varepsilon^{\delta})\rangle_\gamma &= - \varepsilon^\tau (\varepsilon^{\theta})^\alpha (\varepsilon^{\delta})^\beta \Bigg[ G_1(Q^2)~ (2p+q)_\tau ~g_{\alpha\beta}  + G_2(Q^2)~ ( g_{\tau\beta}~ q_\alpha -  g_{\tau\alpha}~ q_\beta) \nonumber\\ &- \frac{1}{2 m_{T_{cc}^+}^2} G_3(Q^2)~ (2p+q)_\tau ~q_\alpha q_\beta  \Bigg],
\end{align}
where $\varepsilon^\delta$ and $\varepsilon^{\theta}$ are the polarization vectors of the initial and final $T_{cc}^+$ states and $\varepsilon^\tau$ is the polarization vector of the photon.  Here, $G_1(Q^2)$, $G_2(Q^2)$ and $G_3(Q^2)$ are invariant form factors,  with  $Q^2=-q^2$.

Using Eqs. (\ref{edmn04})-(\ref{edmn06}), the correlation function takes the form,
\begin{align}
\label{edmn09}
 \Pi_{\mu\nu}^{Had-T_{cc}^+}(p,q) &=  \frac{\varepsilon_\rho \lambda_{T_{cc}^+}^2}{ [m_{T_{cc}^+}^2 - (p+q)^2][m_{T_{cc}^+}^2 - p^2]}
 \Bigg[ G_2 (Q^2) \Bigg(q_\mu g_{\rho\nu} - q_\nu g_{\rho\mu} -
\frac{p_\nu}{m_{T_{cc}^+}^2}  \big(q_\mu p_\rho - \frac{1}{2}
Q^2 g_{\mu\rho}\big) 
 + \nonumber\\
 &  +
\frac{(p+q)_\mu}{m_{T_{cc}^+}^2}  \big(q_\nu (p+q)_\rho+ \frac{1}{2}
Q^2 g_{\nu\rho}\big)
-  
\frac{(p+q)_\mu p_\nu p_\rho}{m_{T_{cc}^+}^4} \, Q^2
\Bigg)
\nonumber\\
&
+\mbox{other independent structures}\Bigg]\,+\cdots.
\end{align}

To determine the MDM, the value of the form factor $G_2(Q^2)$ is needed only at $Q^2 = 0$.
The magnetic form factor $F_M(Q^2)$ is written as 
\begin{align}
\label{edmn07}
&F_M(Q^2) = G_2(Q^2)\,,
\end{align}
 the value of which at $Q^2 = 0 $,  i.e. $F_M(0)$, is proportional to the
 MDM, $\mu_{T_{cc}^+}$ :
\begin{align}
\label{edmn08}
&\mu_{T_{cc}^+} = \frac{ e}{2\, m_{T_{cc}^+}} \,F_M(0).
\end{align}
In  QCD side, the correlation function in Eq. (\ref{edmn01}), is calculated in deep Euclidean region in terms of QCD degrees of freedom as well as the  DAs of the photon. To this end,   we substitute the explicit forms of the  interpolating currents in the correlation function and  contract the corresponding quark fields with the help of the Wick's theorem. As a result,  we  get
\begin{eqnarray}
\Pi _{\mu \nu }^{\mathrm{QCD-T_{cc}^+}}(p,q)&=&i\int d^{4}xe^{ip\cdot x} \langle 0 \mid \Big\{  \mathrm{%
Tr}[ \gamma _{5}\widetilde{S}_{d}^{b^{\prime }b}(-x)\gamma
_{5}S_{u}^{a^{\prime }a}(-x)]    
\mathrm{Tr}[ \gamma _{\nu }\widetilde{S}_{c}^{aa^{\prime
}}(x)\gamma _{\mu }S_{c}^{bb^{\prime }}(x)]\notag \\
&& -\mathrm{Tr}[ \gamma
_{5}\widetilde{S}_{d}^{b^{\prime }b}(-x) \gamma _{5}S_{u}^{a^{\prime }a}(-x)] 
\mathrm{Tr}%
[ \gamma _{\nu }\widetilde{S}_{c}^{ba^{\prime }}(x)\gamma _{\mu
}S_{c}^{ab^{\prime }}(x)] \Big\} \mid 0 \rangle_{\gamma} ,  \label{eq:QCDSide}
\end{eqnarray}%
 where
\begin{equation*}
\widetilde{S}_{c(q)}^{ij}(x)=CS_{c(q)}^{ij\mathrm{T}}(x)C,
\end{equation*}%
with $S_{q(c)}(x)$ being the full light (charm) quark propagator. They are given as
\begin{align}
\label{edmn12}
S_{q}(x)&=i \frac{{\xslash}}{2\pi ^{2}x^{4}} 
- \frac{\langle \bar qq \rangle }{12} \Big(1-i\frac{m_{q} \xslash}{4}   \Big)
- \frac{ \langle \bar qq \rangle }{192}m_0^2 x^2  \Big(1-i\frac{m_{q} \xslash}{6}   \Big)
-\frac {i g_s }{32 \pi^2 x^2} ~G^{\mu \nu} (x) \Big[\rlap/{x} 
\sigma_{\mu \nu} +  \sigma_{\mu \nu} \rlap/{x}
 \Big],
\end{align}
\begin{align}
\label{edmn13}
S_{c}(x)&=\frac{m_{c}^{2}}{4 \pi^{2}} \Bigg[ \frac{K_{1}\Big(m_{c}\sqrt{-x^{2}}\Big) }{\sqrt{-x^{2}}}
+i\frac{{\xslash}~K_{2}\Big( m_{c}\sqrt{-x^{2}}\Big)}
{(\sqrt{-x^{2}})^{2}}\Bigg]
-\frac{g_{s}m_{c}}{16\pi ^{2}} \int_0^1 dv\, G^{\mu \nu }(vx)\Bigg[ \big(\sigma _{\mu \nu }{\xslash}
  +{\xslash}\sigma _{\mu \nu }\big)\nonumber\\
  &\times \frac{K_{1}\Big( m_{c}\sqrt{-x^{2}}\Big) }{\sqrt{-x^{2}}}
+2\sigma_{\mu \nu }K_{0}\Big( m_{c}\sqrt{-x^{2}}\Big)\Bigg],
\end{align}%
where $\langle \bar qq \rangle$ is quark  condensate, $m_0$ is defined through the quark-gluon mixed condensate  $\langle 0 \mid \bar  q\, g_s\, \sigma_{\alpha\beta}\, G^{\alpha\beta}\, q \mid 0 \rangle = m_0^2 \,\langle \bar qq \rangle $, $G^{\mu\nu}$ is the gluon field strength tensor,  $\sigma_{\mu\nu}= \frac{i}{2}[\gamma_\mu, \gamma_\nu]$ and $K_i$'s are modified Bessel functions of the second kind.

Finally, we choose the Lorentz invariant structure $q_\mu \varepsilon_\nu$  from the both sides and match its coefficients from both the hadronic and QCD sides. To eliminate the contributions of the higher states and continuum, we perform Borel transformation and continuum subtraction. The procedures are lengthy but standard, we do not present the steps here and refer the reader for instance to Ref. \cite{Azizi:2018duk}. As a results, we obtain the light-cone QCD sum rule for the MDM as
\begin{align}
 &\mu_{T_{cc}^+}\,\, \lambda_{T_{cc}^+}^2  = e^{\frac{m_{T_{cc}^+}^2}{M^2}} \,\, \Pi^{QCD-T_{cc}^+}(M^2,s_0),
\end{align}
where $M^2$ and $s_0$ are auxiliary parameters stemming from the applications of the Borel transformation and continuum procedures. The $\Pi^{QCD-T_{cc}^+}(M^2,s_0)$  function is quite lengthy,  explicit expression of which  is not presented here.

\subsection{MDM of the\texorpdfstring{$Z_{V}^{++}$}{} state}

As we mentioned in the previous subsection, calculations are started by writing the appropriate correlation function, 
\begin{equation}
 \label{ZV1}
\Pi _{\mu \nu }^{Z_{V}^{++}}(p,q)=i\int d^{4}xe^{ip\cdot x}\langle 0|\mathcal{T}\{J_{\mu}^{Z_{V}^{++}}(x)
J_{\nu }^{Z_{V}^{++}\dagger }(0)\}|0\rangle_{\gamma}, 
\end{equation}%
where 
the $J_{\mu}^{Z_{V}^{++}}(x)$ is interpolating current of the $Z_V^{++}$ state and it is given in the diquark-antidiquark picture with the quantum number $J^P =1^-$ as 
\begin{equation}
J_{\mu }^{Z_V}(x)=\varepsilon \widetilde{\varepsilon }[u_{b}^{T}(x)C\gamma
_{5}c_{c}(x)][\overline{s}_{d}(x)\gamma _{\mu }\gamma _{5}C\overline{d}%
_{e}^{T}(x)],  \label{eq:ZV2}
\end{equation}%
where $\varepsilon \widetilde{\varepsilon }=\varepsilon ^{abc}\varepsilon
^{dec}$, and $a$, $b$, $c$, $d$ and $e$ denote quark colors. In Eq.\ (\ref%
{eq:ZV2}), $u(x)$, $c(x)$, $s(x)$ and $d(x)$ are the quark fields, and $C$ stands for the charge-conjugation operator.

To obtain required sum rules for parameters of $Z_{V}^{++}$ state, we
have to represent the correlation function $\Pi_{\mu\nu}^{Z_{V}^{++}}(p,q)$ with respect to these parameters, and get the hadronic side of
the sum rules $\Pi_{\mu\nu}^{Had-Z_{V}^{++}}(p,q)$. In connection with the hadronic parameters the correlation function has the subsequent form
\begin{align}
\label{ZV3}
\Pi_{\mu\nu}^{Had-Z_{V}^{++}} (p,q) = {\frac{\langle 0 \mid J_\mu^{{Z_{V}^{++}}} \mid
{Z_V^{++}}(p) \rangle}{p^2 - m_{{Z_V^{++}}}^2}} \langle {Z_V^{++}}(p) \mid {Z_V^{++}}(p+q) \rangle_\gamma
\frac{\langle {Z_V^{++}}(p+q) \mid {J^\dagger}_\nu^{{Z_V^{++}}} \mid 0 \rangle}{(p+q)^2 - m_{{Z_V^{++}}}^2} + \cdots,
\end{align}
where the dots stand for contributions of higher resonances and continuum states. The matrix element of the electromagnetic current between the $Z_V^{++}$ states is described by three form factors and it is presented in Eq. (\ref{edmn06}). 
%
%
Using the corresponding equations, the result for the hadronic side is obtained as follows
\begin{align}
\label{ZV5}
 \Pi_{\mu\nu}^{Had-Z_V^{++}}(p,q) &= \frac{\varepsilon_\rho \lambda_{Z_V^{++}}^2}{ [m_{Z_V^{++}}^2 - (p+q)^2][m_{Z_V^{++}}^2 - p^2]}
 \Bigg[ G_2 (Q^2) \Bigg(q_\mu g_{\rho\nu} - q_\nu g_{\rho\mu} -
\frac{p_\nu}{m_{Z_V^{++}}^2}  \big(q_\mu p_\rho - \frac{1}{2}
Q^2 g_{\mu\rho}\big) 
 + \nonumber\\
 &  +
\frac{(p+q)_\mu}{m_{Z_V^{++}}^2}  \big(q_\nu (p+q)_\rho+ \frac{1}{2}
Q^2 g_{\nu\rho}\big)
-  
\frac{(p+q)_\mu p_\nu p_\rho}{m_{Z_V^{++}}^4} \, Q^2
\Bigg)
\nonumber\\
&
+\mbox{other independent structures}\Bigg]\,+\cdots.
\end{align}

The QCD side of the sum rules, $\Pi _{\mu \nu }^{\mathrm{QCD-Z_{V}^{++}}}(p,q)$, is obtained by inserting the interpolating current $J_\mu(x)$ into Eq. (\ref{ZV1}), and contracting proper quark fields via Wick's theorem.  After these steps, we obtain
\begin{eqnarray}
&&\Pi _{\mu \nu }^{\mathrm{QCD-Z_{V}^{++}}}(p,q)=i\int d^{4}xe^{ipx}\varepsilon
\widetilde{\varepsilon }\varepsilon ^{\prime }\widetilde{\varepsilon }%
^{\prime }\langle0\mid \mathrm{Tr}[ \gamma _{5}\widetilde{S}_{u}^{bb^{\prime
}}(x)\gamma _{5}S_{c}^{cc^{\prime }}(x)]   \mathrm{Tr}[ \gamma _{\mu }\gamma _{5}\widetilde{S}%
_{d}^{e^{\prime }e}(-x)\gamma _{5}\gamma _{\nu }S_{s}^{d^{\prime }d}(-x)%
]\mid 0\rangle_{\gamma}.  \label{eq:ZV8}
\end{eqnarray}

The sum rules for the MDM of the $Z_V^{++}$ state can be obtained  by matching the  hadronic and QCD sides of the corresponding correlation function and performing usual operations required in QCD sum rule calculations. As a result of these processes, we get

\begin{align}
 &\mu_{Z_{V}^{++}}\,\, \lambda_{Z_{V}^{++}}^2  = e^{\frac{m_{Z_{V}^{++}}^2}{M^2}} \,\, \Pi^{QCD-{Z_V^{++}}}(M^2,s_0).
\end{align}

The analytical calculations are ended here. Now, we are ready to move on the numerical analysis part.

\end{widetext}

\section{Numerical analysis}

In this section, we perform the numerical analysis of the sum rules for the MDMs of the $T^+_{cc}$ and $Z_V^{++}$  states obtained in the previous section.
We use $m_u=m_d=0$, $m_s =96^{+8}_{-4}\,\mbox{MeV}$,
$m_c = (1.275\pm 0.025)\,$GeV,   $m_{T_{cc}^+}= 3868 \pm 124~\mbox{MeV}$~\cite{Agaev:2021vur},   $m_{Z_{V}^{++}}= 3515 \pm 125~\mbox{MeV}$~\cite{Agaev:2021jsz}
$\langle \bar ss\rangle $= $0.8 \langle \bar uu\rangle$ with 
$\langle \bar uu\rangle $=$(-0.24\pm0.01)^3\,$GeV$^3$~\cite{Ioffe:2005ym},  
$m_0^{2} = 0.8 \pm 0.1$~GeV$^2$~\cite{Ioffe:2005ym}, $\langle \frac{\alpha_s}{\pi} G^2 \rangle =(0.012\pm0.004)$ $~\mathrm{GeV}^4 $~\cite{Belyaev:1982cd}, 
$\lambda_{T_{cc}^+-Di}=(1.96 \pm 0.44)\times 10^{-2}$~GeV$^5$~\cite{Agaev:2021vur},  $\lambda_{T_{cc}^+-Mol}=(1.56 \pm 0.22)\times 10^{-2}$~GeV$^5$ \cite{Xin:2021wcr} and $\lambda_{Z_{V}^{++}}=(1.85 \pm 0.45)\times 10^{-2}$~GeV$^5$~\cite{Agaev:2021jsz}.
The wavefunctions inside the DAs of the photon and all the related parameters are borrowed from Ref.~\cite{Ball:2002ps}.

The next step in the numerical analysis is to find the working regions of the auxiliary parameters known as the continuum threshold $s_0$ and the Borel parameter $M^2$. To this end, we enforce the following conditions: Convergence of operator product expansion, pole dominance, as well as  minimal sensitivity of the resulting observables to auxiliary  parameters. As a result of these constraints, the following working regions are obtained for these parameters:
\begin{align*}
&4.0~\mbox{GeV}^2 \leq M^2 \leq 6.0~\mbox{GeV}^2 ~~ \mbox{for} ~  T_{cc}^+~ \mbox{state},\\
&4.0~\mbox{GeV}^2 \leq M^2 \leq 6.0~\mbox{GeV}^2 ~~ \mbox{for} ~ Z_V^{++}~ \mbox{state},\\
\nonumber\\
&19.5~\mbox{GeV}^2 \leq s_0 \leq 21.5~\mbox{GeV}^2~~ \mbox{for}~ T_{cc}^+~ \mbox{state},\\
&15.0~\mbox{GeV}^2 \leq s_0 \leq 17.0~\mbox{GeV}^2~~\mbox{for}~ Z_V^{++}~ \mbox{state}.
\end{align*}

Having determined the working regions of $M^2$ and $s_0$, we now study the dependence of MDM of the $T^+_{cc}$ and $Z_V^{++}$  states on $M^2$ , at several fixed values of  $s_0$ in Fig. 1. We notice that MDM of the $T^+_{cc}$ and $Z_V^{++}$ states show good stability with respect to the variation in $M^2$ in its working region, especially at lower values of $s_0$. We also find that these MDMs exhibit residual dependence on the variation of $s_0$, which appear as the main sources of uncertainties in the results.
  %
   
  \begin{figure}[htp]
\centering
 \includegraphics[width=0.5\textwidth]{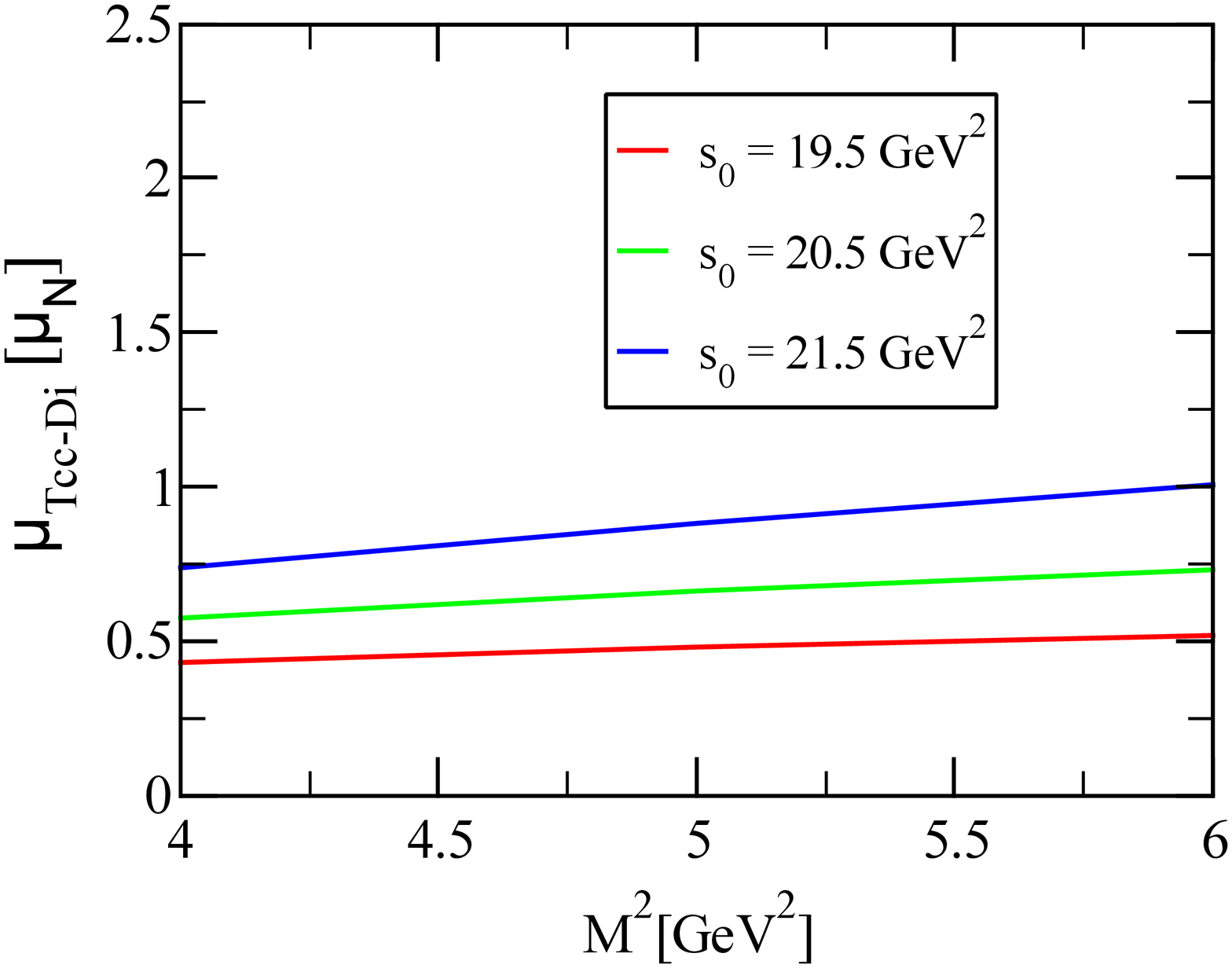}\\
 \includegraphics[width=0.5\textwidth]{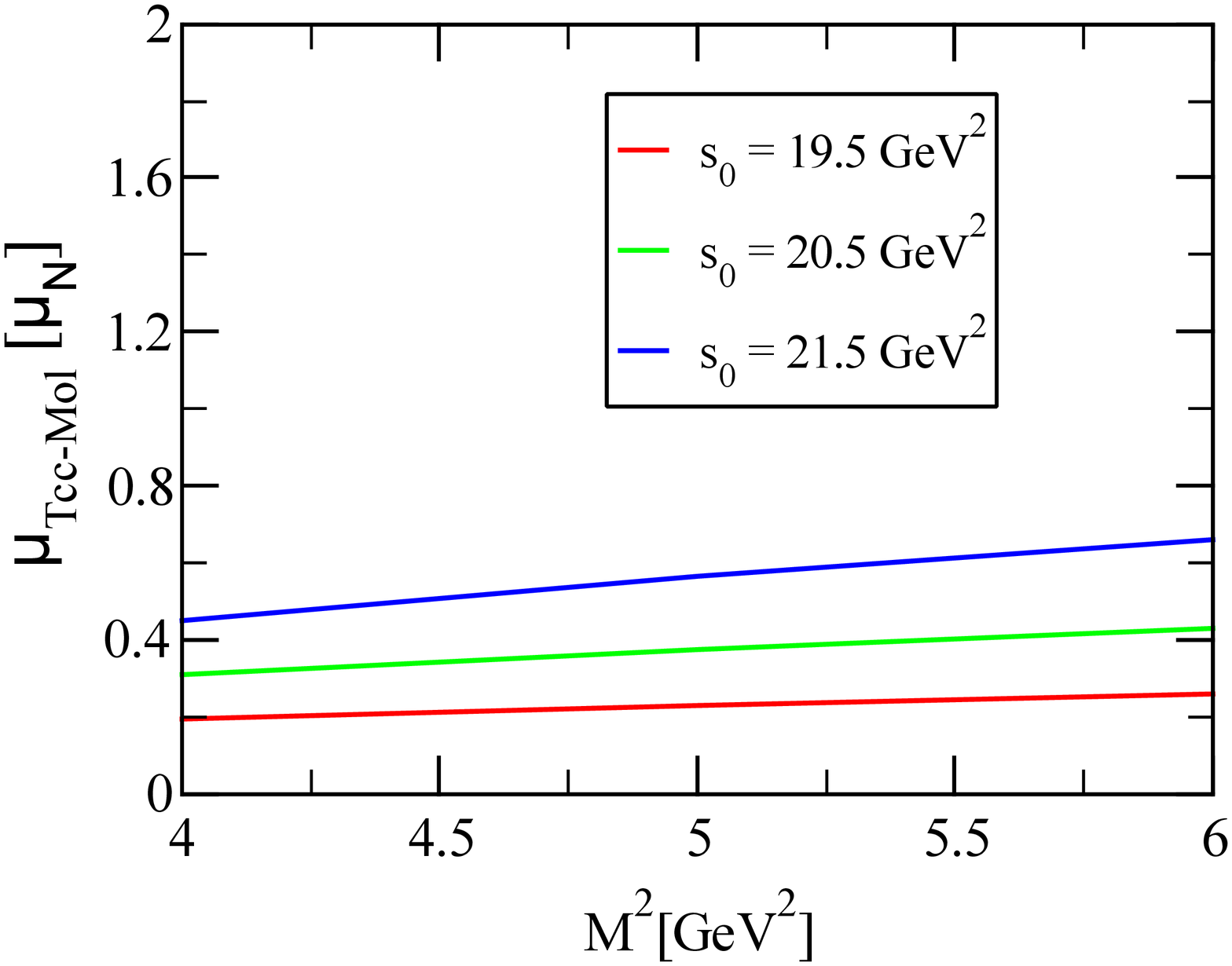}\\
    \includegraphics[width=0.5\textwidth]{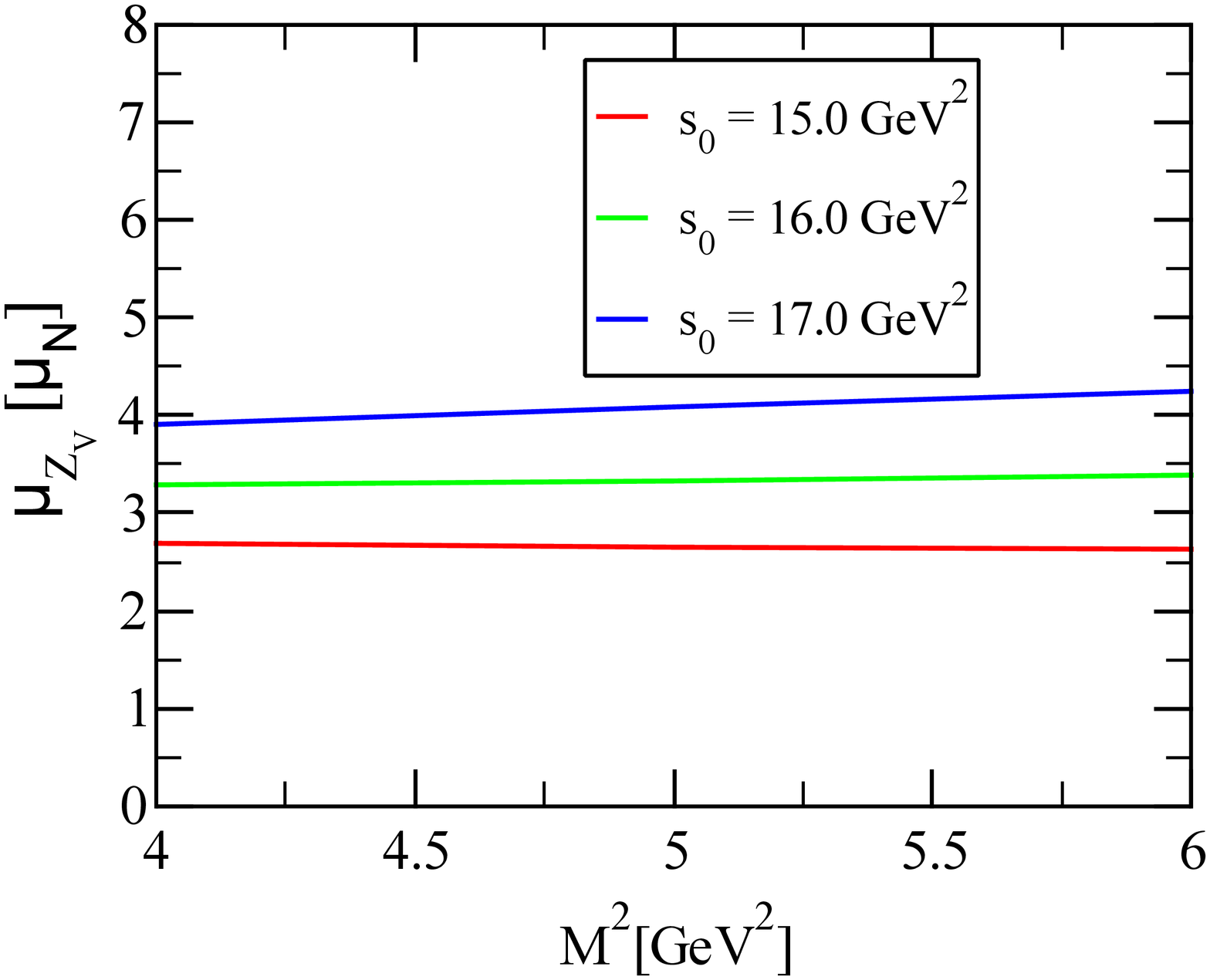}
 \caption{ The dependence of MDMs of the $T_{cc}^+$ and $Z_{V}^{++}$ states on $M^{2}$  at three fixed  values of $s_0$.
 }
  \end{figure}
  

The final  results obtained from the analyses for the MDMs, in both the compact tetraquark of diquark-antidiquark and molecular pictures for  $T_{cc}^+$ state and the compact tetraquark of diquark-antidiquark picture for $Z_{V}^{++}$ state  are given as
\begin{eqnarray}\label{finalresalts}
 &&\mu_{T_{cc}^+-Di} =0.66^{+0.34}_{-0.23}~\mu_N,\\\nonumber
  &&\mu_{T_{cc}^+-Mol} =0.43^{+0.23}_{-0.22}~\mu_N,\\\nonumber
  &&\mu_{Z_{V}^{++}} =3.35^{+0.89}_{-0.73} ~\mu_N.
\end{eqnarray}
Errors in the results are due to the uncertainties related to the calculations of the working regions for $s_0$ and $M^2$ as well as those of other input parameters. It should  be noted  again that the main source of errors in the results is the variations of the results with respect to $s_0$. From Eq. (\ref{finalresalts}), we see that although the central values of the magnetic dipole moments obtained for $T_{cc}^+$ state in two diquark-antidiquark and molecular pictures differ from each other considerably, however they are consistent with each other within the presented uncertainties. This is usual situation in the case of tetraquarks against the standard baryons of three-quarks: the $ \Sigma $ and $ \Lambda $ baryons  contain the same quark content and they are members of the same octet, but they have
very different magnetic dipole moments. This means that the magnetic dipole
moment is not able to distinguish between the two pictures in $T_{cc}^+$  channel from theoretical point of view.  Future possible experimental or lattice results on the magnetic dipole moment of this doubly charmed state and their comparison with the results obtained in the present study as well as comparison of the existing and future experimental results on the mass, width and other decay parameters, related to the various interactions of this state with other particles,  with the theoretical predictions  of different phenomenological models will shed light on the inner structure and nature of this state and help us fix its quantum numbers and quark-gluon organization.  The result for the magnetic dipole moment of $Z_{V}^{++}$ state, due to its double charge, is considerably large compared to other tetraquarks. Any future measurements of its electromagnetic properties and comparison of the results with theoretical predictions  will be very useful in our understanding of tetraquarks and as a result the QCD as theory of strong interaction.

\section{Discussion and concluding remarks}

 The magnetic dipole moments, together with other electromagnetic observables,  of the composite  particles encode important information on their nature, substructure and geometric shape. They contain information on the spatial distributions of charge and magnetization inside  the particles that  directly give information on the spatial distributions of quarks and gluons inside hadrons.

In this study, the MDMs of the recently observed doubly charmed exotic state $T_{cc}^+$ as well as the  recently proposed doubly charged four-quark candidate $Z_V^{++}$  were investigated within the framework of light-cone QCD. We considered both the compact diquark-antidiquark and molecular pictures for $T_{cc}^+$, but only diquark-antidiquark organization for the $Z_V^{++}$ state. They  were treated as states with  $J^P =1^+$ and $J^P =1^-$ quantum numbers, respectively.

 We extracted the values of the MDMs as $\mu_{T_{cc}^+-Di} =0.66^{+0.34}_{-0.23}~\mu_N$, $ \mu_{T_{cc}^+-Mol} =0.43^{+0.23}_{-0.22}~\mu_N $  and $\mu_{Z_{V}^{++}} =3.35^{+0.89}_{-0.73} ~\mu_N$. The obtained results for the $ \mu_{T_{cc}^+} $ in two pictures are consistent within the presented uncertainties although their central values differ from each other considerably. This indicates that the magnetic dipole moment is not able to distinguish between the two pictures. The obtained results in this study for the magnetic dipole moment of this state together with the results of other theoretical models on the mass, width and other decay properties of this state and comparison of the obtained results  with the existing and future experimental data will shed light on the  nature of the newly seen $T_{cc}^+$ tetraquark. The result for the magnetic dipole moment of $Z_{V}^{++}$ state, due to its double charge, is considerably large.  The orders of  MDMs, especially for $\mu_{Z_{V}^{++}}  $,  indicate that they are  accessible in the experiment. Our result on MDM of  $Z_V^{++}$ together with its spectroscopic parameters calculated in Ref. \cite{Agaev:2021jsz}, may help experimental groups in the search for this state and its properties in the experiment. Our results on MDMs may be checked via alternative phenomenological approaches.

\bibliography{article}

\begin{thebibliography}{10}
\expandafter\ifx\csname url\endcsname\relax
  \def\url#1{\texttt{#1}}\fi
\expandafter\ifx\csname urlprefix\endcsname\relax\def\urlprefix{URL }\fi
\expandafter\ifx\csname href\endcsname\relax
  \def\href#1#2{#2} \def\path#1{#1}\fi

\bibitem{Faccini:2012pj}
R.~Faccini, A.~Pilloni, A.~D. Polosa, {Exotic Heavy Quarkonium Spectroscopy: A
  Mini-review}, Mod. Phys. Lett. A 27 (2012) 1230025.
\newblock \href {http://arxiv.org/abs/1209.0107} {\path{arXiv:1209.0107}},
  \href {http://dx.doi.org/10.1142/S021773231230025X}
  {\path{doi:10.1142/S021773231230025X}}.

\bibitem{Esposito:2014rxa}
A.~Esposito, A.~L. Guerrieri, F.~Piccinini, A.~Pilloni, A.~D. Polosa,
  {Four-Quark Hadrons: an Updated Review}, Int. J. Mod. Phys. A 30 (2015)
  1530002.
\newblock \href {http://arxiv.org/abs/1411.5997} {\path{arXiv:1411.5997}},
  \href {http://dx.doi.org/10.1142/S0217751X15300021}
  {\path{doi:10.1142/S0217751X15300021}}.

\bibitem{Chen:2016qju}
H.-X. Chen, W.~Chen, X.~Liu, S.-L. Zhu, {The hidden-charm pentaquark and
  tetraquark states}, Phys. Rept. 639 (2016) 1--121.
\newblock \href {http://arxiv.org/abs/1601.02092} {\path{arXiv:1601.02092}},
  \href {http://dx.doi.org/10.1016/j.physrep.2016.05.004}
  {\path{doi:10.1016/j.physrep.2016.05.004}}.

\bibitem{Ali:2017jda}
A.~Ali, J.~S. Lange, S.~Stone, {Exotics: Heavy Pentaquarks and Tetraquarks},
  Prog. Part. Nucl. Phys. 97 (2017) 123--198.
\newblock \href {http://arxiv.org/abs/1706.00610} {\path{arXiv:1706.00610}},
  \href {http://dx.doi.org/10.1016/j.ppnp.2017.08.003}
  {\path{doi:10.1016/j.ppnp.2017.08.003}}.

\bibitem{Esposito:2016noz}
A.~Esposito, A.~Pilloni, A.~D. Polosa, {Multiquark Resonances}, Phys. Rept. 668
  (2017) 1--97.
\newblock \href {http://arxiv.org/abs/1611.07920} {\path{arXiv:1611.07920}},
  \href {http://dx.doi.org/10.1016/j.physrep.2016.11.002}
  {\path{doi:10.1016/j.physrep.2016.11.002}}.

\bibitem{Olsen:2017bmm}
S.~L. Olsen, T.~Skwarnicki, D.~Zieminska, {Nonstandard heavy mesons and
  baryons: Experimental evidence}, Rev. Mod. Phys. 90~(1) (2018) 015003.
\newblock \href {http://arxiv.org/abs/1708.04012} {\path{arXiv:1708.04012}},
  \href {http://dx.doi.org/10.1103/RevModPhys.90.015003}
  {\path{doi:10.1103/RevModPhys.90.015003}}.

\bibitem{Lebed:2016hpi}
R.~F. Lebed, R.~E. Mitchell, E.~S. Swanson, {Heavy-Quark QCD Exotica}, Prog.
  Part. Nucl. Phys. 93 (2017) 143--194.
\newblock \href {http://arxiv.org/abs/1610.04528} {\path{arXiv:1610.04528}},
  \href {http://dx.doi.org/10.1016/j.ppnp.2016.11.003}
  {\path{doi:10.1016/j.ppnp.2016.11.003}}.

\bibitem{Guo:2017jvc}
F.-K. Guo, C.~Hanhart, U.-G. Mei\ss{}ner, Q.~Wang, Q.~Zhao, B.-S. Zou,
  {Hadronic molecules}, Rev. Mod. Phys. 90~(1) (2018) 015004.
\newblock \href {http://arxiv.org/abs/1705.00141} {\path{arXiv:1705.00141}},
  \href {http://dx.doi.org/10.1103/RevModPhys.90.015004}
  {\path{doi:10.1103/RevModPhys.90.015004}}.

\bibitem{Nielsen:2009uh}
M.~Nielsen, F.~S. Navarra, S.~H. Lee, {New Charmonium States in QCD Sum Rules:
  A Concise Review}, Phys. Rept. 497 (2010) 41--83.
\newblock \href {http://arxiv.org/abs/0911.1958} {\path{arXiv:0911.1958}},
  \href {http://dx.doi.org/10.1016/j.physrep.2010.07.005}
  {\path{doi:10.1016/j.physrep.2010.07.005}}.

\bibitem{Brambilla:2019esw}
N.~Brambilla, S.~Eidelman, C.~Hanhart, A.~Nefediev, C.-P. Shen, C.~E. Thomas,
  A.~Vairo, C.-Z. Yuan, {The $XYZ$ states: experimental and theoretical status
  and perspectives}, Phys. Rept. 873 (2020) 1--154.
\newblock \href {http://arxiv.org/abs/1907.07583} {\path{arXiv:1907.07583}},
  \href {http://dx.doi.org/10.1016/j.physrep.2020.05.001}
  {\path{doi:10.1016/j.physrep.2020.05.001}}.

\bibitem{Liu:2019zoy}
Y.-R. Liu, H.-X. Chen, W.~Chen, X.~Liu, S.-L. Zhu, {Pentaquark and Tetraquark
  states}, Prog. Part. Nucl. Phys. 107 (2019) 237--320.
\newblock \href {http://arxiv.org/abs/1903.11976} {\path{arXiv:1903.11976}},
  \href {http://dx.doi.org/10.1016/j.ppnp.2019.04.003}
  {\path{doi:10.1016/j.ppnp.2019.04.003}}.

\bibitem{Agaev:2020zad}
S.~Agaev, K.~Azizi, H.~Sundu, {Four-quark exotic mesons}, Turk. J. Phys. 44~(2)
  (2020) 95--173.
\newblock \href {http://arxiv.org/abs/2004.12079} {\path{arXiv:2004.12079}},
  \href {http://dx.doi.org/10.3906/fiz-2003-15}
  {\path{doi:10.3906/fiz-2003-15}}.

\bibitem{Dong:2021juy}
X.-K. Dong, F.-K. Guo, B.-S. Zou, {A survey of heavy-antiheavy hadronic
  molecules}, Progr. Phys. 41 (2021) 65--93.
\newblock \href {http://arxiv.org/abs/2101.01021} {\path{arXiv:2101.01021}},
  \href {http://dx.doi.org/10.13725/j.cnki.pip.2021.02.001}
  {\path{doi:10.13725/j.cnki.pip.2021.02.001}}.

\bibitem{LHCb:2021vvq}
R.~Aaij, et~al., {Observation of an exotic narrow doubly charmed
  tetraquark}\href {http://arxiv.org/abs/2109.01038} {\path{arXiv:2109.01038}}.

\bibitem{LHCb:2021auc}
R.~Aaij, et~al., {Study of the doubly charmed tetraquark $T_{cc}^+$}\href
  {http://arxiv.org/abs/2109.01056} {\path{arXiv:2109.01056}}.

\bibitem{Carames:2011zz}
T.~F. Carames, A.~Valcarce, J.~Vijande, {Doubly charmed exotic mesons: A gift
  of nature?}, Phys. Lett. B 699 (2011) 291--295.
\newblock \href {http://dx.doi.org/10.1016/j.physletb.2011.04.023}
  {\path{doi:10.1016/j.physletb.2011.04.023}}.

\bibitem{Richard:2018yrm}
J.-M. Richard, A.~Valcarce, J.~Vijande, {Few-body quark dynamics for doubly
  heavy baryons and tetraquarks}, Phys. Rev. C 97~(3) (2018) 035211.
\newblock \href {http://arxiv.org/abs/1803.06155} {\path{arXiv:1803.06155}},
  \href {http://dx.doi.org/10.1103/PhysRevC.97.035211}
  {\path{doi:10.1103/PhysRevC.97.035211}}.

\bibitem{Hernandez:2019eox}
E.~Hern\'andez, J.~Vijande, A.~Valcarce, J.-M. Richard, {Spectroscopy, lifetime
  and decay modes of the $T^-_{bb}$ tetraquark}, Phys. Lett. B 800 (2020)
  135073.
\newblock \href {http://arxiv.org/abs/1910.13394} {\path{arXiv:1910.13394}},
  \href {http://dx.doi.org/10.1016/j.physletb.2019.135073}
  {\path{doi:10.1016/j.physletb.2019.135073}}.

\bibitem{Liu:2019stu}
M.-Z. Liu, T.-W. Wu, M.~Pavon~Valderrama, J.-J. Xie, L.-S. Geng, {Heavy-quark
  spin and flavor symmetry partners of the X(3872) revisited: What can we learn
  from the one boson exchange model?}, Phys. Rev. D 99~(9) (2019) 094018.
\newblock \href {http://arxiv.org/abs/1902.03044} {\path{arXiv:1902.03044}},
  \href {http://dx.doi.org/10.1103/PhysRevD.99.094018}
  {\path{doi:10.1103/PhysRevD.99.094018}}.

\bibitem{Agaev:2021vur}
S.~S. Agaev, K.~Azizi, H.~Sundu, {Newly observed exotic doubly charmed meson
  $T^{+}_{cc}$}\href {http://arxiv.org/abs/2108.00188}
  {\path{arXiv:2108.00188}}.

\bibitem{Li:2021zbw}
N.~Li, Z.-F. Sun, X.~Liu, S.-L. Zhu, {Perfect $DD^*$ molecular prediction
  matching the $T_{cc}$ observation at LHCb}, Chin. Phys. Lett. 38 (2021)
  092001.
\newblock \href {http://arxiv.org/abs/2107.13748} {\path{arXiv:2107.13748}},
  \href {http://dx.doi.org/10.1088/0256-307X/38/9/092001}
  {\path{doi:10.1088/0256-307X/38/9/092001}}.

\bibitem{Yan:2021wdl}
M.-J. Yan, M.~P. Valderrama, {Subleading contributions to the decay width of
  the $T_{cc}^+$ tetraquark}\href {http://arxiv.org/abs/2108.04785}
  {\path{arXiv:2108.04785}}.

\bibitem{Dong:2021bvy}
X.-K. Dong, F.-K. Guo, B.-S. Zou, {A survey of heavy-heavy hadronic
  molecules}\href {http://arxiv.org/abs/2108.02673} {\path{arXiv:2108.02673}}.

\bibitem{Feijoo:2021ppq}
A.~Feijoo, W.~t. Liang, E.~Oset, {$D^0 D^0 \pi^+$ mass distribution in the
  production of the $T_{cc}$ exotic state}\href
  {http://arxiv.org/abs/2108.02730} {\path{arXiv:2108.02730}}.

\bibitem{Meng:2021jnw}
L.~Meng, G.-J. Wang, B.~Wang, S.-L. Zhu, {Probing the long-range structure of
  the $T_{cc}^+$ with the strong and electromagnetic decays}\href
  {http://arxiv.org/abs/2107.14784} {\path{arXiv:2107.14784}}.

\bibitem{Ling:2021bir}
X.-Z. Ling, M.-Z. Liu, L.-S. Geng, E.~Wang, J.-J. Xie, {Can we understand the
  decay width of the $T_{cc}^+$ state?}\href {http://arxiv.org/abs/2108.00947}
  {\path{arXiv:2108.00947}}.

\bibitem{Xin:2021wcr}
Q.~Xin, Z.-G. Wang, {Analysis of the axialvector doubly-charmed tetraquark
  molecular states with the QCD sum rules}\href
  {http://arxiv.org/abs/2108.12597} {\path{arXiv:2108.12597}}.

\bibitem{Agaev:2020mqq}
S.~S. Agaev, K.~Azizi, B.~Barsbay, H.~Sundu, {Semileptonic and nonleptonic
  decays of the axial-vector tetraquark $T_{bb;\overline{u}
  \overline{d}}^{-}$}, Eur. Phys. J. A 57~(3) (2021) 106.
\newblock \href {http://arxiv.org/abs/2008.02049} {\path{arXiv:2008.02049}},
  \href {http://dx.doi.org/10.1140/epja/s10050-021-00428-5}
  {\path{doi:10.1140/epja/s10050-021-00428-5}}.

\bibitem{Agaev:2020zag}
S.~S. Agaev, K.~Azizi, B.~Barsbay, H.~Sundu, {A family of double-beauty
  tetraquarks: Axial-vector state $T_{bb;\bar{u}\bar{s}}^{-}$}, Chin. Phys. C
  45~(1) (2021) 013105.
\newblock \href {http://arxiv.org/abs/2002.04553} {\path{arXiv:2002.04553}},
  \href {http://dx.doi.org/10.1088/1674-1137/abc16d}
  {\path{doi:10.1088/1674-1137/abc16d}}.

\bibitem{Agaev:2020dba}
S.~S. Agaev, K.~Azizi, B.~Barsbay, H.~Sundu, {Stable scalar tetraquark
  $T_{bb;\bar{u}\bar{d}}^{-}$}\href {http://arxiv.org/abs/2001.01446}
  {\path{arXiv:2001.01446}}, \href
  {http://dx.doi.org/10.1140/epja/s10050-020-00187-9}
  {\path{doi:10.1140/epja/s10050-020-00187-9}}.

\bibitem{Agaev:2019lwh}
S.~S. Agaev, K.~Azizi, B.~Barsbay, H.~Sundu, {Heavy exotic scalar meson
  $T_{bb;\bar{u}\bar{s}}^{-}$}, Phys. Rev. D 101~(9) (2020) 094026.
\newblock \href {http://arxiv.org/abs/1912.07656} {\path{arXiv:1912.07656}},
  \href {http://dx.doi.org/10.1103/PhysRevD.101.094026}
  {\path{doi:10.1103/PhysRevD.101.094026}}.

\bibitem{Agaev:2019kkz}
S.~S. Agaev, K.~Azizi, H.~Sundu, {Double-heavy axial-vector tetraquark
  $T_{bc;\bar{u}\bar{d}}^{0}$}, Nucl. Phys. B 951 (2020) 114890.
\newblock \href {http://arxiv.org/abs/1905.07591} {\path{arXiv:1905.07591}},
  \href {http://dx.doi.org/10.1016/j.nuclphysb.2019.114890}
  {\path{doi:10.1016/j.nuclphysb.2019.114890}}.

\bibitem{Agaev:2019qqn}
S.~S. Agaev, K.~Azizi, H.~Sundu, {Strong decays of double-charmed pseudoscalar
  and scalar $cc\overline{u}\overline{d}$ tetraquarks}, Phys. Rev. D 99~(11)
  (2019) 114016.
\newblock \href {http://arxiv.org/abs/1903.11975} {\path{arXiv:1903.11975}},
  \href {http://dx.doi.org/10.1103/PhysRevD.99.114016}
  {\path{doi:10.1103/PhysRevD.99.114016}}.

\bibitem{Agaev:2018khe}
S.~S. Agaev, K.~Azizi, B.~Barsbay, H.~Sundu, {Weak decays of the axial-vector
  tetraquark $T_{bb;\bar{u} \bar{d}}^{-}$}, Phys. Rev. D 99~(3) (2019) 033002.
\newblock \href {http://arxiv.org/abs/1809.07791} {\path{arXiv:1809.07791}},
  \href {http://dx.doi.org/10.1103/PhysRevD.99.033002}
  {\path{doi:10.1103/PhysRevD.99.033002}}.

\bibitem{Agaev:2018vag}
S.~S. Agaev, K.~Azizi, B.~Barsbay, H.~Sundu, {The doubly charmed pseudoscalar
  tetraquarks $T_{cc;\bar{s} \bar{s}}^{++}$ and $T_{cc;\bar{d} \bar{s}}^{++}$},
  Nucl. Phys. B 939 (2019) 130--144.
\newblock \href {http://arxiv.org/abs/1806.04447} {\path{arXiv:1806.04447}},
  \href {http://dx.doi.org/10.1016/j.nuclphysb.2018.12.021}
  {\path{doi:10.1016/j.nuclphysb.2018.12.021}}.

\bibitem{Aaij:2020hon}
R.~Aaij, et~al., {A model-independent study of resonant structure in $B^+\to
  D^+D^-K^+$ decays}, Phys. Rev. Lett. 125 (2020) 242001.
\newblock \href {http://arxiv.org/abs/2009.00025} {\path{arXiv:2009.00025}},
  \href {http://dx.doi.org/10.1103/PhysRevLett.125.242001}
  {\path{doi:10.1103/PhysRevLett.125.242001}}.

\bibitem{Aaij:2020ypa}
R.~Aaij, et~al., {Amplitude analysis of the $B^+\to D^+D^-K^+$ decay}, Phys.
  Rev. D 102 (2020) 112003.
\newblock \href {http://arxiv.org/abs/2009.00026} {\path{arXiv:2009.00026}},
  \href {http://dx.doi.org/10.1103/PhysRevD.102.112003}
  {\path{doi:10.1103/PhysRevD.102.112003}}.

\bibitem{Burns:2020xne}
T.~J. Burns, E.~S. Swanson, {Discriminating among interpretations for the
  $X(2900)$ states}, Phys. Rev. D 103~(1) (2021) 014004.
\newblock \href {http://arxiv.org/abs/2009.05352} {\path{arXiv:2009.05352}},
  \href {http://dx.doi.org/10.1103/PhysRevD.103.014004}
  {\path{doi:10.1103/PhysRevD.103.014004}}.

\bibitem{Agaev:2021jsz}
S.~S. Agaev, K.~Azizi, H.~Sundu, {Doubly charged vector tetraquark
  $Z_V^{++}$=$[cu][\bar s \bar d]$}, Phys. Lett. B 820 (2021) 136530.
\newblock \href {http://arxiv.org/abs/2105.00081} {\path{arXiv:2105.00081}},
  \href {http://dx.doi.org/10.1016/j.physletb.2021.136530}
  {\path{doi:10.1016/j.physletb.2021.136530}}.

\bibitem{Ozdem:2021yvo}
U.~\"Ozdem, K.~Azizi, {Magnetic dipole moment of the $Z_{cs}(3985)$ state:
  diquark-antidiquark and molecular pictures}, Eur. Phys. J. Plus 136 (2021)
  968.
\newblock \href {http://arxiv.org/abs/2102.09231} {\path{arXiv:2102.09231}},
  \href {http://dx.doi.org/10.1140/epjp/s13360-021-01977-w}
  {\path{doi:10.1140/epjp/s13360-021-01977-w}}.

\bibitem{Brodsky:1992px}
S.~J. Brodsky, J.~R. Hiller, {Universal properties of the electromagnetic
  interactions of spin one systems}, Phys. Rev. D 46 (1992) 2141--2149.
\newblock \href {http://dx.doi.org/10.1103/PhysRevD.46.2141}
  {\path{doi:10.1103/PhysRevD.46.2141}}.

\bibitem{Azizi:2018duk}
K.~Azizi, A.~R. Olamaei, S.~Rostami, {Beautiful mathematics for beauty-full and
  other multi-heavy hadronic systems}, Eur. Phys. J. A 54~(9) (2018) 162.
\newblock \href {http://arxiv.org/abs/1801.06789} {\path{arXiv:1801.06789}},
  \href {http://dx.doi.org/10.1140/epja/i2018-12595-1}
  {\path{doi:10.1140/epja/i2018-12595-1}}.

\bibitem{Ioffe:2005ym}
B.~L. Ioffe, {QCD at low energies}, Prog. Part. Nucl. Phys. 56 (2006) 232--277.
\newblock \href {http://arxiv.org/abs/hep-ph/0502148}
  {\path{arXiv:hep-ph/0502148}}, \href
  {http://dx.doi.org/10.1016/j.ppnp.2005.05.001}
  {\path{doi:10.1016/j.ppnp.2005.05.001}}.

\bibitem{Belyaev:1982cd}
V.~M. Belyaev, B.~L. Ioffe, {Determination of the baryon mass and baryon
  resonances from the quantum-chromodynamics sum rule. Strange baryons}, Sov.
  Phys. JETP 57 (1983) 716--721.

\bibitem{Ball:2002ps}
P.~Ball, V.~M. Braun, N.~Kivel, {Photon distribution amplitudes in QCD}, Nucl.
  Phys. B 649 (2003) 263--296.
\newblock \href {http://arxiv.org/abs/hep-ph/0207307}
  {\path{arXiv:hep-ph/0207307}}, \href
  {http://dx.doi.org/10.1016/S0550-3213(02)01017-9}
  {\path{doi:10.1016/S0550-3213(02)01017-9}}.

\end{thebibliography}

\end{document}